# Accurate Prediction of Chemical Shifts for Aqueous Protein Structure for "Real World" Cases using Machine Learning

Jie Li[1,2], Kochise C. Bennett[1,2], Yuchen Liu[1,2], Michael V. Martin[3], Teresa Head-Gordon[1-4*]

[1]Pitzer Center for Theoretical Chemistry, [2]Department of Chemistry, [3]Department of Bioengineering, [4]Department of Chemical and Biomolecular Engineering
University of California, Berkeley CA 94720

Accurate prediction of NMR chemical shifts can in principle help refine aqueous solution structure of proteins to the quality of X-ray structures. We report a new machine learning algorithm for protein chemical shift prediction that outperforms existing chemical shift calculators on realistic NMR solution data. Our UCBShift predictor implements two modules: a transfer prediction module that employs both sequence and structural alignment to select reference candidates for experimental chemical shift replication, and a redesigned machine learning module based on random forest regression which utilizes more, and more carefully curated, feature extracted data. When combined together, this new predictor achieves state-of-the-art accuracy for predicting chemical shifts on a "real-world" dataset, with root-mean-square errors of 0.31 ppm for amide hydrogens, 0.19 ppm for Hα, 0.87 ppm for C', 0.81 ppm for Cα, 1.01 ppm for Cβ, and 1.83 ppm for N, exceeding current prediction accuracy of popular chemical shift predictors such as SPARTA+ and SHIFTX2.

**INTRODUCTION**

Chemical shifts are a readily obtainable NMR observable that can be measured with high accuracy for proteins, and sensitively probe the local electronic environment that can yield quantitative information about protein secondary structure[1-3], estimation of backbone torsion angles[4,5], or measuring the exposure of the amino acid residues to solvent[6]. But in order to take full advantage of these high quality NMR measurements, there is a necessary reliance on a computational model that can relate the experimentally measured NMR shifts to structure with high accuracy. Existing methods for chemical shift prediction rely on extensive experimental databases together with useful heuristics to rapidly, but non-rigorously, simulate protein chemical shifts. As of yet, quantum mechanical methods which would in principle provide more rigor to chemical shift prediction are still in progress[7].

The heuristic chemical shift back-calculators are formulated as either alignment-based methods such as SHIFTY[8] and SPARTA[9,10], or feature-based methods including SHIFTCALC[11], SHIFTX[12], PROSHIFT[13], CamShift[14] and SPARTA+[15], or in the case of SHIFTX2[16], both. The alignment-based methods rely on the fact that proteins with similar sequences will also share similar structures which lead to their exhibiting similar chemical shifts. This idea was first exploited by the program SHIFTY[8] which "transferred" the chemical shifts from known sequences in the database to the query sequence based on a global sequence alignment. Higher accuracy was achieved in the formulation of SHIFTY+ by replacing the global sequence alignment with a local sequence alignment, and is included in the most recent chemical shift prediction program SHIFTX2[16]. Alignment-based methods in general yield predictions with higher accuracy when a good sequence homologue is found in the database, and the constant increase in the number of sequences and associated chemical shifts deposited into the Biomolecular Magnetic Resonance Bank (BMRB)[17], suggest that a similar sequence to the query sequence will continue to increase steadily.

On the other hand, methods that are based on sequence alignments will by definition fail if sequence similarity between the query sequence with any sequence in the database is too low, as well as the possibility that similar sequences can adopt very different structural folds[18]. For query sequences with low sequence identity, the feature extraction methods predict secondary chemical shifts (i.e. from a random coil reference[19]) by providing data input formulated as hypersurfaces of structural data attributes such as backbone ϕ, ψ angles and hydrogen bonding derived from X-ray structures or calculated from quantum mechanics, or physical data observables such as ring currents[20,21], electric field effects[22], or Lipari-Szabo order parameters[23], that can be generated from easily parsed computational models. These feature extracted data are then used to establish empirical hypersurfaces such as used in SHIFTS[7,24] and CamShift[14], or to train a machine learning algorithm in the cases of PROSHIFT[13], SPARTA+[15], and the SHIFTX+ component of SHIFTX2[16].

We have developed and tested a new generation algorithm, UCBShift, for solution chemical shift prediction for all relevant protein atom types, with root mean square errors (RMSEs) of 0.31, 0.19, 0.87, 0.81, 1.01 and 1.83 ppm for H, Hα, C', Cα, Cβ and N respectively when evaluated on any independently generated test sequence. UCBShift utilizes small molecule information (water and ligands), physically inspired non-linear transformation of features derived from structure, together with a two-level machine learning pipeline that exploits sequence as well as structural alignments, to achieve this current state-of-the-art performance.

## METHODS

**Preparation of Training and New Testing Datasets.** A high-quality database of protein structures and associated accurately referenced chemical shifts are crucial for composing a machine learning approach that can make reliable predictions of the chemical shifts, and for faithfully comparing the performance of existing alternative approaches such as SPARTA+ and SHIFTX2. Several publicly available data sources, including the SPARTA+ training set and the training and testing set for SHIFTX+, were combined into a single training dataset that captures the structure and chemical shift relationship. Since all of these data were used in the development of the original SPARTA+ and SHIFTX2 methods, it ensures that corrections for chemical shift reference values were included in our dataset as well.

Unlike previous incarnations of these data sets, which stripped out all presence of crystal waters and ligands, we generated a data set that retained the small molecules in the crystal structures. Our hypotheses is that for crystal waters especially, they often are highly conserved and functional, and are likely highly populated even in solution NMR experiments.[25,26]. Any reported hydrogen atoms in the Research Collaboratory for Structural Bioinformatics protein databank (RCSB or PDB) structures[27] were removed and a systematic approach for adding a complete set of hydrogens used the program REDUCE to keep consistency in the structural data used across all approaches. To ensure more robust training, for each atom type, residues with chemical shifts deviating from the average by 5 standard deviations and residues that DSSP[28] failed to generate secondary structure predictions were removed, which accounted for the removal of 40, 5, 18, 147 and 1 training examples for H, Hα, Cα, Cβ and N shifts, respectively. When stereochemically inequivalent Hα were present, their shifts were averaged. In the creation of data for each of the individual atom types, any residues that do not have recorded chemical shifts in RefDB are eliminated.

Before excluding redundant chains from the database, there were altogether 852 proteins in the training dataset. Duplicate chains were identified and excluded from our dataset: two chains are regarded as duplicates if the sequences and their structures are exactly the same, or eliminating the shorter sequence if it is a sub-sequence of a longer sequence (which is kept). However, 32 chains in the SPARTA+ dataset were retained because although they had identical sequences, they were found to have different structures and thus different chemical shifts. After excluding the duplicate chains by this prescription, the number of protein chains in the training dataset decreased to 647. The filtering of the training dataset based on RCI-S[229] in principle excludes flexible residues whose chemical shifts are harder to predict. We did not exclude training data based on RCI-S[2] as was done in some other chemical shift predictors, because a complete training set that covers different prediction difficulties is crucial for obtaining reliable performance for real-world applications. Table 1 reports the total number of training data examples for each of the different atom types. In addition, the RefDB database[30], which is a database for re-referenced protein chemical shifts assignments extracted and corrected from BMRB, was also compiled for the alignment-based chemical shifts prediction.

Since the training dataset in Table 1 covers all of the data from SPARTA+ and SHIFTX2, a separate test dataset needed to be prepared for a fair comparison of all of the chemical shift programs. Therefore, 200 proteins with high-resolution (<2.4 Å) X-ray structures and with chemical shifts available in the RefDB were selected at random to form a separate test set that do not share the exact same sequence as training structures. These structures were downloaded from RCSB and again hydrogens were added with the REDUCE algorithm. Erroneous chemical shifts assignments were removed from this test dataset, which include 9 (H, Cβ, and N) chemical shifts that were significantly offset from the random coil average (Table S1); 8 Cβ chemical shifts from cysteines that show strong disagreement with their expected chemical shifts under their oxidation state in the crystal structures (Table S2), and all C' chemical shifts from 3 proteins that are anti-correlated with predictions from SPARTA+, SHIFTX2 and UCBShift (Figure S1).

**Table 1.** Total number of training and testing examples for chemical shift prediction for each atom type. The data eliminates duplicate chains, and residues with no deposited chemical shift values.

|  | # of PDBs | H | Hα | C' | Cα | Cβ | N |
|---|---|---|---|---|---|---|---|
| **Train** | 647 | 72894 | 56149 | 58228 | 79611 | 70621 | 74896 |
| **Test** | 200 | 19120 | 11727 | 8231 | 13140 | 10139 | 15374 |
| **Test (Curated)** | 200 | 18494 | 11240 | 7861 | 12533 | 9883 | 14610 |
| **LH-Test** | 100 | 8634 | 4979 | 3332 | 5685 | 4278 | 6576 |
| **LH-Test (Curated)** | 100 | 8606 | 4950 | 3331 | 5251 | 4201 | 6480 |

It is essential to remove these chemical shifts from the test set because of evidence for the existence of experimental or recording errors in these data; but no further processing was done on the test set so that it is a good representation of a "real-world" application. A more carefully "curated" test dataset based on these 200 proteins was also prepared, which additionally exclude paramagnetic proteins, some Hα chemical shifts that have calculated ring current effect exceeding 1.5 ppm, "outliers" detected by the PANAV program[10], and chemical shifts corrected by PANAV that are different from their original values by more than 0.3 ppm for hydrogens, 1.0 ppm for carbons and 1.5 ppm for nitrogen. These additional test filters are similar to the procedures used by SPARTA+ and SHIFTX2 in preparing their test datasets[15,16]. A complete list of the 200 testing proteins are given in the supporting information (Tables S3). As is inevitable, some of these 200 proteins share high sequence identity with some of the training data, so we also generate test datasets after filtering out proteins with >30% sequence identity to yield a low-homology testset (LH-Test) with 100 test proteins (Table 1).

**Machine Learning for Chemical Shift Prediction.** The new UCBShift chemical shift prediction program is composed of two sub-modules: the transfer prediction module (UCBShift-Y) that utilizes sequence and structural alignments to "transfer" the experimental chemical shift value to the query example, and a machine learning module (UCBShift-X) that learns the mapping between the feature extracted data to the experimental chemical shift in the training data. The overall structure of the algorithm is depicted in Figure 1.

*UCBShift-Y module.* UCBShift-Y is similar in spirit to the SHIFTY+ component of SHIFTX2, in that the experimental chemical shift for a given atom type in a given residue can be transferred to the query protein when the sequence of the protein in the database is highly similar or even identical to the sequence of the query protein. However, instead of relying on the sequence alignment alone, we have developed an algorithm that relies on both sequence alignment and structural alignment, which allows for proteins that are highly related in structure but remotely related in sequence to be utilized. The use of structural alignments also prevents proteins that have high sequence similarity but low structural homology to mislead an algorithm to erroneous chemical shifts transfers.

For the UCBShift-Y module, a query sequence is first aligned with all sequences in the RefDB database using the local BLAST algorithm[31], and the PDB files for all sequences generating significant matches are further aligned with the query PDB structure using the mTM-align algorithm[32]. The alignments were further filtered to only keep those alignments that have TM score greater than 0.5 and an RMSD with the query protein structure that is smaller than 2.8 Å. For each of the aligned PDB sequences, its best alignment with the refDB sequence is determined using the Needleman-Wunsch alignment[33].

If the residues are exactly the same, the shifts from RefDB are directly transferred to the target; otherwise, the secondary chemical shifts from RefDB are transferred to account for the different chemical shift reference states for different amino acids. To be more specific, the target shift for atom A and residue I is calculated from the matching residue J:

$$\delta_{I,A} = \delta_{rc,I,A} + (\delta_{J,A} - \delta_{rc,J,A}) \tag{1}$$

where $\delta_{rc,I,A}$ and $\delta_{rc,J,A}$ are the random coil shifts for atom type A in residue I and J, respectively, and $\delta_{J,A}$ is the chemical shift for the matching residue in the database.

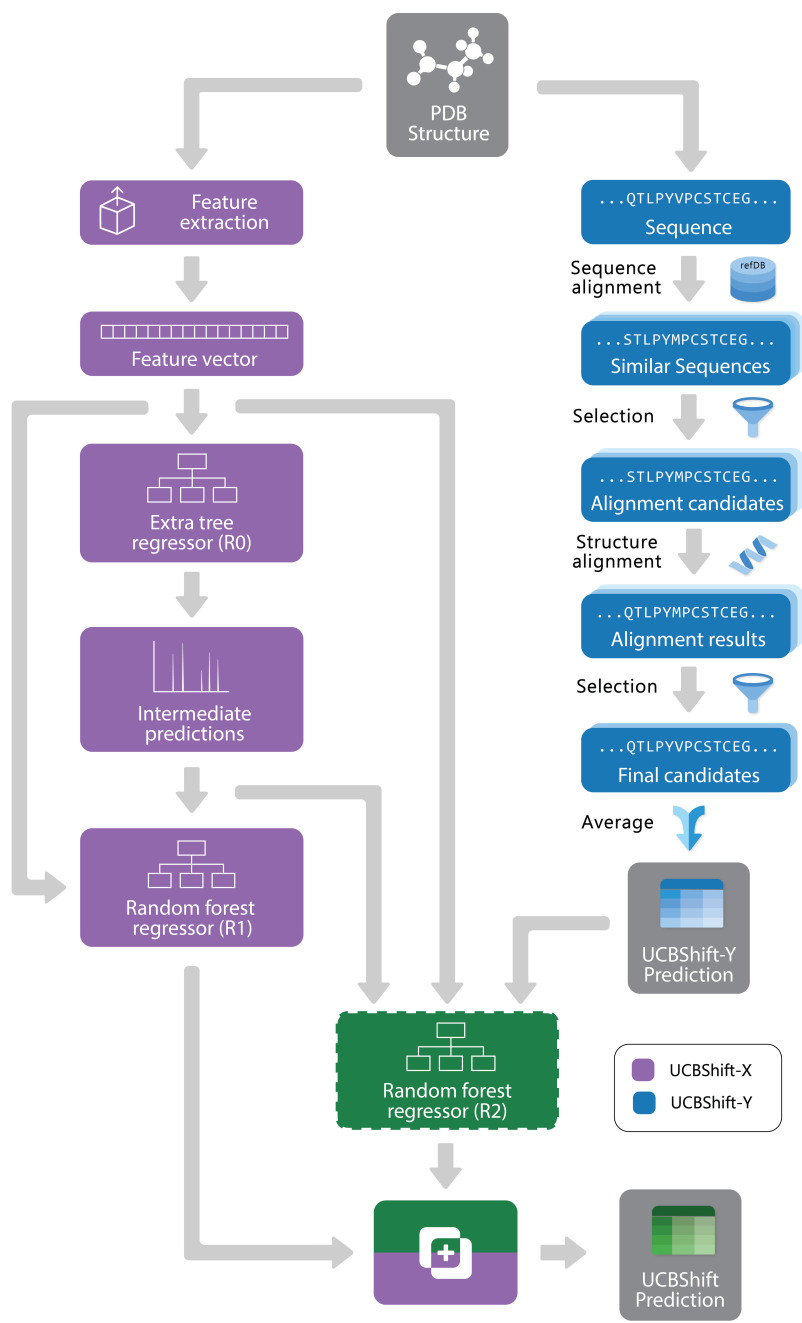

**Figure 1**. *The overall design of the UCBShift chemical shift prediction algorithm*. It combines both a transfer prediction module that relies on both sequence and structural alignments, and a machine learning module that trains a tree regression model on augmented feature extracted data.

When multiple significant structural alignments exist for a given residue, the secondary shifts from these references are averaged with an exponential weighting $w_I$,

$$w_I = e^{5*(S_{NA} \times S_{TM})+B_{IJ}} * \mathbf{1}(B_{IJ} \geq 0) \qquad (2)$$

given by the normalized sequence alignment score, $S_{NA} = S_{blast}/\max(S_{blast})$, where $S_{blast}$ is the blast score of the matching sequence, and $\max(S_{blast})$ is the maximum blast score from all blast hits; the structure alignment TM score $S_{TM}$ is the pairwise TM score between the query structure and the aligned structure, and $B_{IJ}$ is the substitution likelihood between the residue in the query sequence and the residue in the matching sequence using the BLOSUM62 substitution matrix[34]. Weights with negative substitution scores are set to zero.

*UCBShift-X module*. The UCBShift-X module requires the formulation of feature extracted data of a given atom type in a query residue, and the ability to map the feature extracted data to the chemical shift value during the training. Similar to the SPARTA+ program or for the SHIFTX+ component of SHIFTX2, we have developed residue-specific features for the query residue and the previous and next residues to the query residue, but we have included more features and polynomial transformations of the features to improve prediction. The feature extracted data generated from the PDB structures of individual residues include

- 20 numbers representing the score for substituting the residue to any other amino acid, and taken from the BLOSUM62 substitution matrix[34]
- sine and cosine values of the ϕ and ψ dihedral angles at the residue. Taking the sine and cosine values of the dihedral angles prevents the discontinuity when the dihedral angle goes from +180° to -180°. For the undefined dihedral angles, for example the ϕ angle of a residue at the N-terminus and the ψ angle of a residue at the C-terminus, both the sine and cosine values were set to zero.
- a binary number indicating whether $\chi_1$ or $\chi_2$ dihedral angles for the side chain exists (existence indicator), and the sine and cosine values of these angles when they are defined for the same reasons described for backbone dihedral angles.
- existence indicators and geometric descriptors for the hydrogen bond between the amide hydrogen and carboxyl oxygen, and between the Cα hydrogen and a carboxyl group (so called α-hydrogen bonds). For each position in the query residue that hydrogen bonds can form, a group of five numbers describe the properties of the hydrogen bond: a Boolean number indicating its existence, the distance between the closest hydrogen bond donor-acceptor pair, the cosine values for the angles at the donor hydrogen atom and at the acceptor atom, and the energy of the hydrogen bond calculated with the DSSP model[28].

For the query residue, all hydrogen descriptors for amide hydrogen, carboxyl oxygen and α hydrogen are included, but only the carboxyl oxygen features are included for the previous residue, and the amide hydrogen features for the next residue. These add up to 25 hydrogen bond descriptors for any given residue.

- $S^2$ order parameters calculated by the contact model[23]
- absolute and relative atomic surface area produced by the DSSP program
- hydrophobicity of the residue by the Wimley-White whole residue hydrophobicity scales[35].
- ring current effect calculated by the Haigh-Mallion model[20,21]. For each training model for a specific atom type, the ring current for that atom type is included, while the ring currents for other atom types are excluded from the feature set.
- the one-hot representation of the secondary structure of the residue produced by DSSP program (composed of eight categories)
- average B factor of the residue extracted from the PDB file
- half-sphere exposure of the residue[36]
- Polynomial transformations of some of the residue-specific features, such as the hydrogen bond distances ($d_{HB}$), by including $d_{HB}^2$, $d_{HB}^{-1}$, $d_{HB}^{-2}$, $d_{HB}^{-3}$, and the squares of the cosine values of the dihedral angles are also included as additional features. These polynomial quantities have been found to be correlated with secondary chemical shifts, and have occurred in several empirical formulas for calculating chemical shifts.[3,37]

Unlike SPARTA+ and SHIFTX+, we have developed a pipeline with an extra tree regressor[38] followed by random forest regressor[39] as the machine learning based predictor shown in Figure 1. Both the extra tree regressor and random forest regressor are ensembles of tree regressors that split the data using a subset of the features, and make ensemble-based predictions via a majority vote. However, extra tree regressors split the nodes in each tree randomly by selecting an optimal cut-point from uniformly distributed cut-points in the range of the feature, while the random forest regressors calculate the locally optimal cut in a feature by comparing the information entropy difference before and after the split. The random forest regressor learns based on the predictions from the first tree regressor and all the other input features, which can be regarded as a variant of the boosting algorithm[40], since it learns from the mistakes the first predictor makes. The pipeline was first optimized using the TPOT tool[41] with 3-fold cross validation on the training set, and all the parameters were fine-tuned using a temporal validation dataset with 50 structures randomly selected from the training set. Because tree-based ensemble models are robust to the inclusion of irrelevant features[42], feature selection was not performed. A more detailed analysis of the feature importance will be given in the Results.

Algorithmically, two separate random forest (RF) regressors are trained. The first RF regressor (R1) only accepts features extracted from the structure and the prediction from the extra tree regressor, and the second RF regressor (R2) additionally takes the secondary shift output from UCBShift-Y, together with the maximal and average identity for sequence matches in the database, and is trained using only a subset of the training data for which UCBShift-Y is able to make a prediction. Based on the availability of UCBShift-Y predictions, the final prediction of the whole algorithm is generated either by R1 (when no UCBShift-Y predictions are available) or R2 (when UCBShift-Y is able to make predictions). Finally, the random coil reference values are added back to the prediction to complete the total chemical shift prediction, i.e. the predictions are calculated with

$$\delta_{pred} = \begin{cases} f_{R1}(X, f_{R0}(X)) + \delta_{RC} & \text{when } UCBShift-Y \text{ generates no prediction} \\ f_{R2}(X, f_{R0}(X), S) + \delta_{RC} & \text{when } UCBShift-Y \text{ generates predictions} \end{cases} \quad (3)$$

where $f_{R0}$ represents the first-level extra tree regressor, $f_{R1}$ and $f_{R2}$ are the two second-level random forest regressors, X are all the features extracted from the structure, S are the predictions from UCBShift-Y and the identity scores, and $\delta_{RC}$ is the random coil chemical shift for the given residue.

**RESULTS**

The performance of SPARTA+, SHIFTX2, and UCBShift are evaluated across the newly created test dataset of 200 proteins (Test) and the subset of 100 low sequence homology with respect to the training set (LH-Test), each of which is uncurated or curated as described in Methods (Table 2). In general, the performance of SPARTA+ is even across both the curated Test and curated LH-Test datasets. The average RMSE error for SPARTA+ predictions on the uncurated Test and LH-Test datasets increases further, in which we provide the minimum error and the maximum error for each protein in graphical form in Figures S2 and S3.

SHIFTX2 is seen to outperform SPARTA+ on all atom types on the curated dataset when there is high sequence homology for which it was designed, and it performs comparably to SPARTA+ on the curated data for target sequences with low sequence similarity to the training data. However, we find that the actual performance on curated data set is less accurate than the reported performance of the SHIFTX2 method [16]. One possible explanation is that a sequence similarity analysis revealed that out of the original 61 testing proteins of SHIFTX2, 4 proteins had 100% sequence alignment with a protein in the training dataset, sometimes under different identification numbers (Table S4). This problem of training data leakage into the testing data of the original SHIFTX2 method could be a non-trivial source of the better performance of SHIFTX2 reported in the literature. The protein-specific average RMSE error and the scatter plots for the SHIFTX2 predicted and experimental shifts are also given in Figure S2 and S3 on the uncurated Test dataset.

**Table 2.** *Test performance of SPARTA+, SHIFTX2, and UCBShift for chemical shift predictions of relevant atom types found in proteins.* We compare the performance of SPARTA+[(a)], SHIFTX2[(b)], and UCBShift across an independently generated uncurated test dataset of 200 proteins that do not share the same sequence as the training set (Test) and a subset of 100 proteins with < 30% sequence identity to the training set (LH-Test). We also compare the 3 methods against filtered test data that removes post-prediction chemical shift data outliers. All in units of ppm.

| Dataset | Test | | | LH-Test | | |
|---|---|---|---|---|---|---|
| Atom Type | SPARTA+ | SHIFTX2 | UCBShift | SPARTA+ | SHIFTX2 | UCBShift |
| H | 0.51 | 0.44 | **0.31** | 0.49 | 0.49 | **0.45** |
| Hα | 0.27 | 0.23 | **0.19** | 0.27 | 0.26 | **0.26** |
| C' | 1.25 | 1.16 | **0.87** | 1.16 | 1.20 | **1.14** |
| Cα | 1.16 | 1.05 | **0.81** | 1.13 | 1.15 | **1.09** |
| Cβ | 1.35 | 1.27 | **1.01** | 1.36 | 1.37 | **1.34** |
| N | 2.72 | 2.40 | **1.83** | 2.73 | 2.73 | **2.61** |
| Dataset | Test (Curated) | | | LH-Test (Curated) | | |
| Atom Type | SPARTA+ | SHIFTX2 | UCBShift | SPARTA+ | SHIFTX2 | UCBShift |
| H | 0.49 | 0.42 | **0.30** | 0.48 | 0.47 | **0.43** |
| Hα | 0.26 | 0.22 | **0.18** | 0.26 | 0.25 | **0.24** |
| C' | 1.15 | 1.06 | **0.79** | 1.16 | 1.19 | **1.13** |
| Cα | 1.09 | 0.98 | **0.76** | 1.08 | 1.10 | **1.04** |
| Cβ | 1.17 | 1.09 | **0.83** | 1.15 | 1.15 | **1.12** |
| N | 2.59 | 2.25 | **1.73** | 2.67 | 2.66 | **2.54** |

(a) reported SPARTA+ values from [15]: 0.49 ppm for H, 0.25 ppm for Hα, 1.09 ppm for C', 0.94 ppm for Cα, 1.14 ppm for Cβ, and 2.45 ppm for N.
(b) reported SHIFTX2 values from [43]: 0.17 ppm for H, 0.12 ppm for Hα, 0.53 ppm for C', 0.44 ppm for Cα, 0.52 ppm for Cβ, and 1.12 ppm for N.

By comparison we find that filtering of the test set for outliers that disagree with the predictions, the elimination of paramagnetic proteins, and removing test shifts for hydrogen due to potentially inaccurate and large ring currents effects has more limited effect on prediction performance. To illustrate, the distributions of absolute errors from SPARTA+ for paramagnetic proteins and diamagnetic proteins in the Test dataset are shown in Figure S4. The error distributions are not that different for H, Hα, Cβ, and N, and while paramagnetic proteins show higher prediction errors than diamagnetic proteins for the C' and Cα data types, they are not egregious errors.

We find that UCBShift outperforms SPARTA+ and the SHIFTX2 algorithm when tested on the uncurated test data set, the more carefully curated test data, and regardless of the level of sequence homology. The protein-specific average RMSE error and the scatter plots for the UCBShift predicted and experimental shifts are also given in Figure S2 and S3 on the uncurated Test dataset. In order to understand the improved performance of UCBShift in particular, we analyze the components of the algorithm including the UCBShift-X and UCBShift-Y modules, as well as the importance of the extracted features, utilizing the full test set of 200 proteins in more detail below.

**Analysis of UCBShift-Y module**. The major difference of our transfer prediction module in comparison with SHIFTY or SHIFTY+ is the inclusion of a structural alignment to select reference sequences for subsequent sequence alignment. There is a trade-off between the coverage the transfer learning module can achieve and the average accuracy of the prediction, requiring tuning the thresholds for accepting an imperfectly aligned protein as reference. Empirically we chose a relatively permissive threshold for sequence alignment to enable sequences that do not have much similarity with the query protein proceed to the next step in case it generates a good structure alignment. The thresholds for the TM score and RMSD in structure alignment were optimized during training to ensure the reference structures are close enough to the query structure. A TM score threshold of 0.5 was selected because usually structures that have TM score greater than 0.5 have roughly the same fold.

We hypothesized that a structure-based alignment followed by sequence alignment would be more reliable since it would (1) allow for transferring shifts from structurally homologous proteins with low sequence identity, while also (2) ensuring that the transferred chemical shift values are not from a protein that has high sequence similarity but low structural homology with the query protein. This is confirmed in Figure 2 which plots the difference of the RMSE on amide hydrogen chemical shift prediction between our UCBShift-Y and SHIFTY+ as a function of sequence identity. Here the sequence identity is defined as the ratio of the number of matched residues to either the length of the query sequence or the length of the matched sequence, whichever is longer. Furthermore, the UCBShift results are reported with a specifically designed "test mode" which will not utilize sequences with more than 99% identity with the query sequence for making the prediction; this practice ensures the testing performance is a more realistic reflection of the actual performance when operating on input data which is not included in the search database. It is evident that on average predictions on query sequences with low sequence similarity but high structural homology are greatly improved with UCBShift-Y.

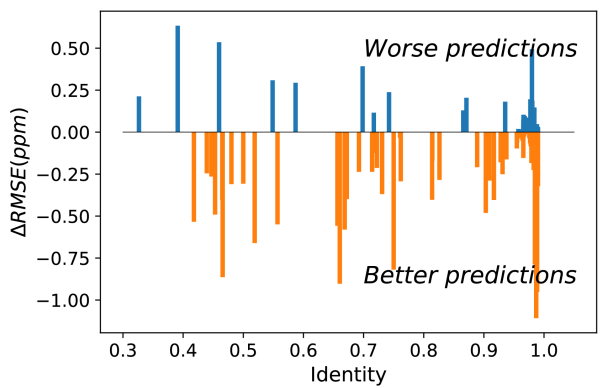

**Figure 2.** *Difference between UCBShift-Y and SHIFTY+ for protein specific RMSEs for amide hydrogens as a function of sequence identity.* The presence of more negative values indicates UCBShift-Y is making better predictions than SHIFTY+ across the range of sequence identity.

A particularly interesting example is the prediction for adenylate kinase (PDB ID: 4AKE)[44] and its mutant (PDB ID: 1E4V)[45], both of which are identical but for a single substitution of a valine for a glycine residue at position 10 (Figure 3). Even with such high sequence identity, these two proteins adopt quite different tertiary structures with a backbone RMSD of 7.08 Å as can be seen from the overlay of their two structures in Figure 3a. Hence while the experimental chemical shifts for these two proteins have a root-mean-square difference (RMSD) of 0.38 ppm for amide hydrogen shifts overall, the maximum H chemical shift difference is much larger at 1.34 ppm and is reflected in the surprisingly lower correlation (R-value=0.86) between the amide hydrogen shifts for two proteins given the high sequence similarity (Figure 3b). Therefore when using SHIFTY or SHIFTY+ for the 1E4V query sequence, the best sequence match will be 4AKE, thus increasing the chemical shift prediction error due to the huge structural deviation between the two proteins. Instead when using our UCBShift-Y module it selects two alternative proteins, 1AKE and 2CDN, which share an average sequence similarity of only 67% with the query protein. The correlation between the predicted 1E4V amide hydrogen shifts with UCBShift-Y which chooses references based on structural alignment and the experimental values are given in Figure 3c, raising the R-value to 0.94 and lowering the RMSE to 0.25 ppm.

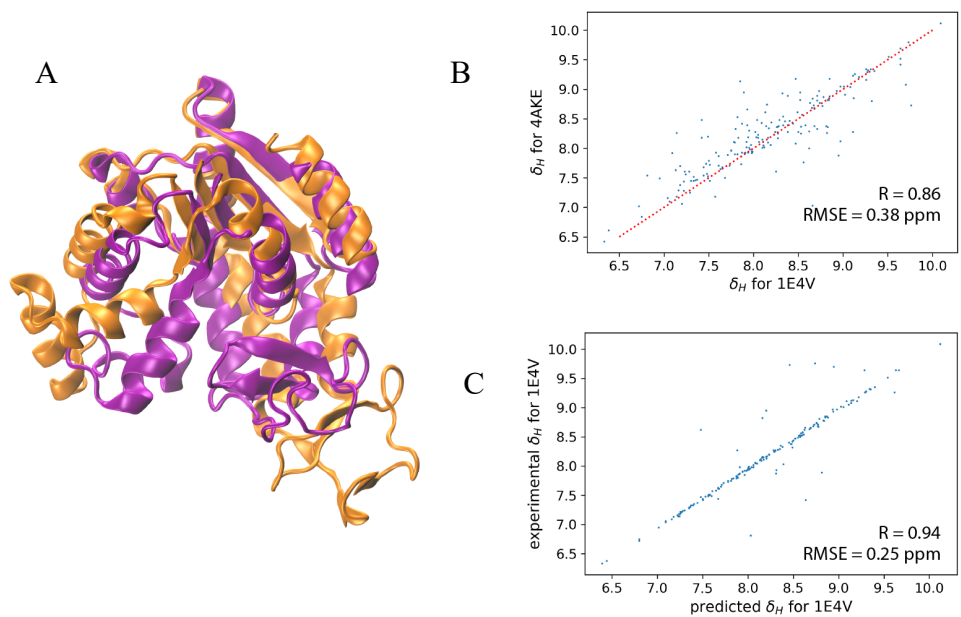

**Figure 3**. *Analysis of the transfer prediction module for UCBShift which uses sequence and structural alignment.* (a) Structural alignment of adenylate kinase (4AKE, orange) and the mutant G10V of adenylate kinase (1E4V, purple). (b) Correlation between experimental chemical shifts of the amide hydrogen for 4AKE and 1E4V. (c) Correlation between predicted amide hydrogen chemical shifts using UCBShift with experimental values. In this case structural alignments instead of sequence alignments were used for selecting references for the transfer prediction.

Figure S5 summarizes the results of UCBShift-Y vs. SHIFTY+ for chemical shift prediction for all atom types, validating that the structural alignments successfully found better reference proteins for the query protein which improved the overall prediction quality. In comparison with SHIFTY+, all atom types other than carboxyl carbon are improved in accuracy; although predictions for the carboxyl carbons are at the same level of accuracy as SHIFTY+, the failure to improve this atom type with UCBShift-Y is likely due to the lower number of chemical shifts available for transfer prediction for this atom type. Finally we note that our UCBShift-Y can be used as a standalone chemical shift predictor when sequence and structural alignments exist and have available experimental chemical shifts.

**Analysis of the UCBShift-X module.** The connection between features extracted from PDB files and the secondary chemical shifts was explored using several machine learning methods, including neural networks with a single hidden layer, deep fully-connected neural networks, residual neural networks, convolutional neural networks, recurrent neural networks, as well as tree-based ensemble models. The more complex and deeper neural networks performed well on the training dataset and validation dataset, however their performance on the test data was found to be no better than SPARTA+ or SHIFTX+, likely indicating that more feature extracted data is needed and/or due to problems with data representation, to fully exploit the potential of these methods. Thus the tree-based ensemble models stood out as a more competitive machine learning predictor for chemical shifts with limited data.

**Table 3.** *RMSE for the individual elements of transfer prediction (UCBShift-Y) and machine learning module (UCBShift-X) on the test dataset.* The standalone UCBShift-Y prediction when sequence and structural alignments exist and have available experimental chemical shifts. The chemical shift prediction of the machine learning module (UCBShift-X) which is trained independent of any transfer prediction. The prediction of the R2 module with input from UCBShift-Y module, and the combined R1+R2 modules that defines the UCBShift calculator.

| UCBShift components | H | Hα | C' | Cα | Cβ | N |
|---|---|---|---|---|---|---|
| **UCBShift-X (R1)** | 0.44 | 0.25 | 1.47 | 1.10 | 1.48 | 2.50 |
| **UCBShift-Y** | 0.22 | 0.20 | 0.66 | 0.56 | 0.77 | 1.69 |
| **ML with UCBShift-Y input (R2)** | 0.21 | 0.16 | 0.76 | 0.59 | 0.75 | 1.34 |
| **UCBShift (R1+R2)** | 0.31 | 0.19 | 1.18 | 0.82 | 1.13 | 1.84 |

The RMSE of the pipeline with extra tree regressor and random forest regressor but without inputs from UCBShift-Y (R1) between the predicted chemical shifts and the observed shifts is summarized in Table 3 and named UCBShift-X. It is found to be better for all the atom types when compared with SPARTA+, or the SHIFTX+ component of SHIFTX2, which also use no sequence and/or structural alignments. The overall performance of the UCBShift-X machine learning module is promising, and it also can be used as a reliable standalone predictor for chemical shifts, especially when no faithful alignment is found using UCBShift-Y.

If we consider using the R2 module (which is trained using only a subset of the training data for which UCBShift-Y is able to make a prediction), the errors of some atom types further decrease (Table 3). Interestingly, the averaged prediction error from R2 for H, Hα, Cβ and N is even smaller than the average error of UCBShift-Y, indicating that the second ML module is doing better than just combining the results from UCBShift-Y and from the first level machine learning module R0 for these atom types. But given the uncertainties in sequence and structural alignments or the lack of chemical shift data for UCBShift-Y, the R1 and R2 machine learning modules are combined to yield the final algorithm and results for chemical shifts as given in Table 3 for all the six atom types.

**Analysis of the Data Representation.** A further test is done to analyze the contributions of different features extracted from the structural PDB files to the R0, R1, and R2 pipelines that define the UCBShift algorithm (Figure 1). Relative feature importance is calculated as the total decrease in node impurity weighted by the probability of reaching a node decided with that feature, and averaged over all the trees in the ensemble.[46] The results are analyzed on the predictions for amide hydrogen as a working example, and are given in Table 4.

**Table 4.** *Importance of different input features into the R0, R1, and R2 pipelines of the machine learning modules.*

| Feature categories | R0 | R1 | R2 |
|---|---|---|---|
| **Backbone dihedral angles** | 0.23 | 0.12 | 0.05 |
| **Transformed features** | 0.23 | 0.11 | 0.04 |
| **Secondary structure** | 0.16 | 0.005 | 0.001 |
| **BLOSUM numbers** | 0.11 | 0.02 | 0.006 |
| **Hydrogen bond** | 0.11 | 0.06 | 0.015 |
| **Half surface exposure** | 0.05 | 0.05 | 0.02 |
| **Ring current** | 0.04 | 0.03 | 0.006 |
| **Sidechain dihedral angles** | 0.03 | 0.03 | 0.005 |
| **Atomic surface area** | 0.02 | 0.01 | 0.002 |
| **B factor** | 0.007 | 0.01 | 0.002 |
| **$S^2$ order parameters** | 0.005 | 0.01 | 0.002 |
| **Hydrophobicity** | 0.002 | 0.001 | 0.0003 |
| **pH values** | 0.001 | 0.002 | 0.001 |
| **Prediction from R0** | N/A | 0.54 | 0.18 |
| **TP prediction and similarity** | N/A | N/A | 0.67 |

For the R0 module we find that the most predictive features are the backbone dihedral angles, the secondary structure, BLOSUM numbers, hydrogen bond features, and the ring current effect which are included in SPARTA+ and SHIFTX2. However, the polynominal transformations of the structural data and half surface exposure are unique in our feature set, and they have relatively high importance among all the features for the R0 component. Not surprisingly, the R0 input is nearly half of the important features for R1, but the backbone dihedral angles and the polynomial transformations are ~25% of the important extracted features. The UCBShift-X prediction and similarity as well as the prediction from R0 are the dominant factors for the R2 model; this result indicates the network is indeed trying to differentiate situations when UCBShift-X predictions are more reliable and when they are not so accurate in comparison to R0 predictions, as well as based on the other structure-derived features. Therefore, using machine learning to combine the predictions from feature-based prediction and alignment-based prediction is a better strategy than doing a weighted average of the two predictions. Finally features such as hydrophobicity and pH values, and to some extent B-factors and $S^2$ order parameters, seem to play a minor role in predictive capacity of the ML module.

## DISCUSSION AND CONCLUSION

Prediction of protein chemical shifts from structure has relied on robust and popular algorithms such as SPARTA+ and SHIFTX2 that represent the 3-dimensional structure by a set of extracted features that are presented to a machine learning algorithm, sometimes supplemented with direct transfer of experimental data taken on related proteins of a given query sequence. In this paper, we tested the performance of SPARTA+ and SHIFTX2 on a large test set of proteins not previously encountered in previous training and test sets, and showed that SPARTA+ performs as reported and evenly across high and low sequence homology test data, as expected. SHIFTX2 still outperforms SPARTA+ on test sequences with high sequence homology, but not at the same levels expected from the reported RMSE literature values[43]. In this work we show that higher accuracy for chemical shifts can be achieved with an enhanced hybrid algorithm, UCBShift, that makes predictions using machine learning on a more extended set of extracted features and transferring experimental chemical shifts from a database by utilizing both sequence and structural alignments. The feature extraction algorithm, the UCBShift prediction program, and all training and testing data can be downloaded from a publicly available github repository https://thglab.berkeley.edu/software-and-data/).

Although the performance of these algorithms are much better when applied to carefully curated test data, the filtering out of test data risks the inability to distinguish between a poor prediction from a poor experimental chemical shift value. Large outliers would certainly result from the wrong random coil reference for Cβ shifts due to ambiguous cysteine oxidation states, or single whole proteins which exhibit many chemical shift outliers for particular atom types, and should not be considered a failure in algorithmic performance, but a problem of the experimental data. However, further test filtering can start to become arbitrary as we move from deviant to suspicious to acceptable experimental agreement with *the prediction*; one can't have it both ways. Thus in this paper we have provided a realistic range of test performance since scientists use these chemical shift predictors on real-world data that may differ from the original training datasets such that the algorithms do not generalize well -i.e. some measure of disagreement with experiment may just simply be prediction error. As such Table 2 provides a more realistic range of test reliability for all three methods.

Although we have realized noticeable improvement over other protein chemical shifts predictors, we believe we are reaching the limit of accuracy by using extracted feature from structures or transfer predictions through alignments. Deep learning may be helpful in the next step since it can operate directly on 3D data representations without the potential bias introduced by features extracted by human experts, as we have shown recently for chemical prediction in the solid state, in which we greatly improved prediction for all atom types and approached chemical accuracy on par with ab initio calculations for hydrogen in particular.[47] The ability to move to 3D representations will be important for the solution NMR chemical shift predictions for intrinsically disordered proteins, since feature extracted data will be less available and likely less representative for this class of protein.

**SUPPLEMENTARY DATA**
- Exclusion of erroneous chemical shifts assignments in the test dataset
- Compilation of test dataset
- Evaluation of SPARTA+ and SHIFTX2 performance on uncurated Test dataset
- Performance analysis of UCBShift-Y in comparison with SHIFTY+

**FUNDING**
This work was supported by the National Institutes of Health [5U01GM121667]; and the computational resources of the National Energy Research Scientific Computing Center, a DOE Office of Science User Facility supported by the Office of Science of the U.S. Department of Energy [DE-AC02-05CH11231].

**ACKNOWLEDGEMENT**
KCB thanks the California Alliance Postdoctoral Fellowship for early support of this work. We also thank Mojtaba Haghighatlari, Brad Ganoe, Tim Stauch, Lars Urban, Shuai Liu, and Martin Head-Gordon for fruitful discussions.

**GRAPHICAL ABSTRACT**

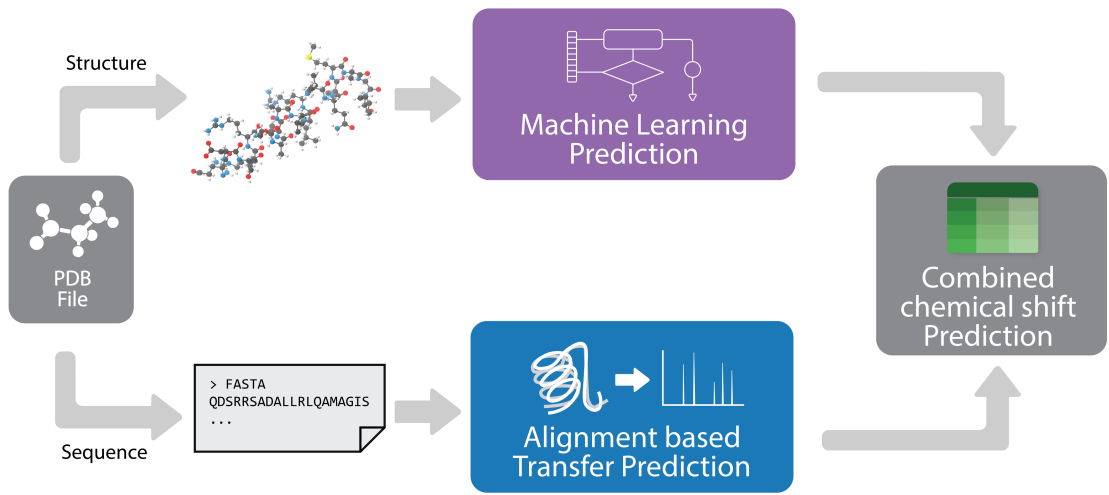

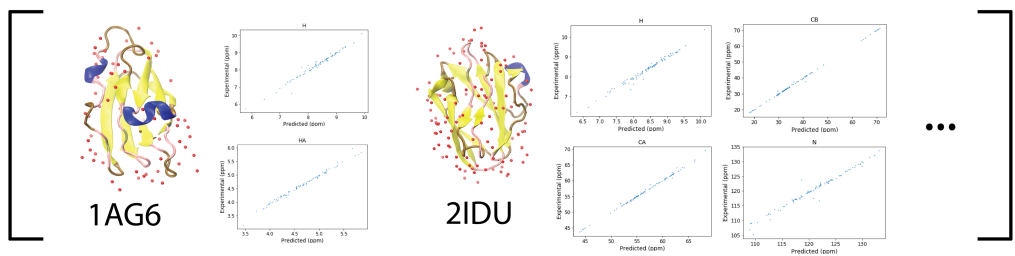

Evaluation on real-world dataset